\documentclass[10pt,aps,twocolumn,nofootinbib]{revtex4-1}
\usepackage{amsfonts}
\usepackage{mathrsfs}
\usepackage{amsmath}
\usepackage{amssymb}
\usepackage[normalem]{ulem}
\usepackage{extarrows}
\usepackage{slashed}
\usepackage{isodateo}
\usepackage{graphicx} 
\usepackage{xcolor}
\usepackage[bookmarksnumbered=true,bookmarksopen=true]{hyperref}
 \hypersetup{colorlinks,%
             linkcolor=[rgb]{0,0.3,0.6}, %
             citecolor=[rgb]{0,0.3,0.6}, %
             urlcolor=[rgb]{0,0.3,0.6}}%
\renewcommand{\rm}{\mathrm} 
\begin{document}

\title{Resolving XENON Excess With Decaying Cold Dark Matter}
\author{Shuai Xu}
\email{shuaixu@cqu.edu.cn}
\affiliation{Department of Physics, Chongqing University, Chongqing 401331, China}
\author{Sibo Zheng}
\email{sibozheng.zju@gmail.com}
\affiliation{Department of Physics, Chongqing University, Chongqing 401331, China}

\begin{abstract}
We propose a decaying cold dark matter model to explain the excess of electron recoil observed at the XENON1T experiment. 
In this scenario, the daughter dark matter from the parent dark matter decay easily obtains velocity large enough to saturate the peak of the electron recoil energy around 2.5 keV,
and the observed signal rate can be fulfilled by the parent dark matter with a mass of order $10-200$ MeV and a lifetime  larger than the age of Universe.
We verify that this model is consistent with experimental limits from dark matter detections, Cosmic Microwave Background and Large Scale Structure experiments.
\end{abstract}
\maketitle

\section{Introduction}
%%%% BackgroundDM
Recently,  XENON1T experiment \cite{Aprile:2020tmw}, which is a dark matter (DM) direct detection facility, 
has reported an excess of electron recoil over the background in the $1-7$ keV range with $3.5\sigma$ significance.
As pointed out in ref.\cite{Aprile:2020tmw}, this excess is unlikely due to solar axion, neutrino magnetic moment or statistical uncertainties about the background.
So far, the observed excess has initiated extensive investigations about potential astrophysical sources.
Among other things, a cold DM is a natural candidate, which is the subject of this study.

To explain the observed excess, one has to accommodate two critical quantities 
- the electron recoil energy range around 2-3 keV and the  electron transfer momentum range near 50 keV.
Unfortunately, they conflict with a conventional cold DM, 
which has a velocity  typically of order $\sim 10^{-3}c$, with $c$ the velocity of light.
A few proposals have been proposed to avoid the conflicts.
In the case of elastic scattering \cite{Kannike:2020agf,Fornal:2020npv,Su:2020zny, Bell:2020bes,Cao:2020bwd,Jho:2020sku},  
the cold DMs can be boosted in certain circumstances before they interact with the electrons in the xenon atoms,
while in the case of inelastic scattering \cite{Harigaya:2020ckz,Bramante:2020zos, Aboubrahim:2020iwb, Lee:2020wmh,Baryakhtar:2020rwy} the favored electron transfer momentum range can be realized  in terms of small  rest mass splitting between two different DM components.

In this study, we propose a new decaying cold DM scenario, 
in which the parent DM (A) decays to the daughter DM particles (B),
\begin{eqnarray}{\label{decay}}
A \rightarrow B B.
 \end{eqnarray}
In terms of the decay, the velocity of the daughter DM can be enhanced to be comparable with $c$, 
relative to the small velocity of the parent DM.
Unlike photoelectric absorption in a decaying warm DM \cite{Choi:2020udy},  
the daughter particle  elastically scatters  off the electrons in the xenon atoms.
We will show that without any violations of current DM (in)direct detections or cosmological measurements,
this decaying cold DM model can easily resolve the XENON excess.

\section{The Model}
We begin with the production of the daughter particle $B$ due to $A$ decay.
The decay yields the following velocity and present number density of $B$ respectively,
\begin{eqnarray}{\label{B}}
\frac{\upsilon_{B}}{c}&=& \sqrt{\frac{m^{2}_{A}}{4m^{2}_{B}}-1},\label{velocity}\\
n_{B} &=& 2\frac{\rho_{\text{dm},0}}{m_{A}}\left[1-\exp{(-\Gamma_{A}\rm{t}_{0})}\right], \label{number}
 \end{eqnarray}
where $m_A$ and $m_{B}$ refer to the parent and daughter DM mass respectively with $m_{A}>2m_{B}$,
$\rm{t}_{0}$ is the age of Universe, 
while $\rho_{\text{dm},0}=0.4~\rm{GeV}/\text{cm}^{3}$ and $\Gamma_{A}$ denote the local DM density and the decay width of the parent DM $A$, respectively.
We will assume that compared to the decay production the thermal production of $n_B$ is subdominant.

The input parameters in Eqs.(\ref{velocity})-(\ref{number}) are constrained as follows. 
Firstly,  in order to yield $\upsilon_{B}$ of order $\sim 0.1~c$ the mass ratio $m_{A}/2m_{B}$ should deviate from unity in percent level,
which implies that $m_A$ and $2m_{B}$ are highly degenerate. 
Secondly, in order to fulfill the cosmological bounds on the decaying DM both from the Cosmic Microwave Background (CMB) and Large Scale Structure (LSS) experiments,
the lifetime of $A$ particle $\tau_{A}=\Gamma^{-1}_{A}$ should be larger than $\rm{t}_{0}$,
which can be achieved by adjusting the coupling constant between $A$ and $B$, 
with the help of a suppression by the small $\beta$ factor due to the mass degeneracy.

We will return to the cosmological constraints after we have explored the signal rate of the recoil electrons at the XENON1T.

\section{Signal Rate}
 According to conservations of energy and momentum in the elastic scattering process, 
the energy\footnote{In the case of inelastic scattering, $E_e$ in Eq.(\ref{EC}) receives a new term due to the rest mass splitting between two different components that involve in the scattering off electron.} transferred to electron reads \cite{Bloch:2020uzh} 
\begin{eqnarray}{\label{EC}}
E_{e}=\mathbf{q}\cdot \overrightarrow{\upsilon}_{B}-\frac{q^{2}}{2m_{B}}.
 \end{eqnarray}
From Eq.(\ref{EC}) the maximal value  $E_{e}^{\rm{max}}\approx \frac{1}{2}m_{B} \upsilon^{2}_{B}$ at  $q\approx m_{B}\upsilon_{B}$ valid only when $m_{B}\approx m_{e}$.
Consider that when $m_{B}\leq m_{e}$ the daughter DM with a large velocity is severely constrained by limits such as the effective number of  neutrinos,
we will focus on $m_{B}\gg m_{e}$, under which $E_{e}^{\rm{max}}\approx 2m_{e}\upsilon^{2}_{B}$ \cite{Kannike:2020agf} instead.
This constraint implies  $\upsilon_{B}\geq  0.05~c$ in order to satisfy $E_{e}^{\rm{max}}\geq 2.5$ keV.

Given a fixed value of $E_e$, Eq.(\ref{EC}) determines the electron transfer momentum range $q_{-}<q<q_{+}$, with 
\begin{eqnarray}{\label{q}}
q_{\pm}=m_{B}\upsilon_{B}\pm \sqrt{m^{2}_{B}\upsilon^{2}_{B}-2m_{B}E_{e}}.
 \end{eqnarray}
The transfer momentum range in Eq.(\ref{q}) affects the signal rate of the recoil electrons discussed below 
in the sense that the atomic factorization factor $K(E_{e},q)$ \cite{Roberts:2019chv, Roberts:2016xfw} is rather sensitive to $q$,
which takes the maximal value $K_{\rm{max}}\approx 0.1$ at $q_{\rm{peak}}\approx 50$ keV for $E_{e}=2$ keV, 
and dramatically declines as $q$ slightly deviates from $q_{\rm{peak}}$.
Therefore, in order to maximize the $K$-factor contribution to the signal rate,
we should take suitable values of $m_{B}$ and $\upsilon_{B}$ to make sure that $q_{\rm{peak}}$ is covered by the electron transfer momentum range in Eq.(\ref{q}).

Furthermore, the daughter DM-free electron scattering cross section $\bar{\sigma}_e$ relies on the nature of mediator \cite{Alhazmi:2020fju}
which communicates the interaction between the daughter DM B and electron.
From the viewpoint of effective field theory, $\bar{\sigma}_e$ can be written as 
\begin{eqnarray}{\label{cs}}
\bar{\sigma}_{e}\approx \frac{g^{2}_{mB}g^{2}_{me}m^{2}_{e}}{\pi m^{4}_{\rm{med}}},
 \end{eqnarray}
 where $m_{\rm{med}}$ is the mediator mass,
$g_{mB}$ is the coupling between the mediator and $B$, 
and $g_{me}$ is the coupling between the mediator and electron.
If the mediator is identified as a standard model particle, only $g_{mB}$ in Eq.(\ref{cs}) is a free parameter,
the magnitude of which has to be constrained by the decay width of the standard model particle.
 
After a handle on the ``luminosity" and the DM B-electron scattering cross section, 
we now estimate the number of events of recoil electrons
\begin{eqnarray}{\label{R}}
\frac{dR}{dE}&\approx& n_{\rm{xe}}n_{B}\times\frac{\bar{\sigma}_{e}}{2m_{e}\upsilon_{B}}\\
&\times& \int dE_{e}\left[\int^{q_{+}}_{q_{-}} dq a^{2}_{0} q \mid F(q) \mid^{2} K(E_{e}, q)\right]R_{s}(E,E_{e}),\nonumber
 \end{eqnarray}
where $n_{\rm{xe}}\approx  4.2\times 10^{27}/$ton is the number density of xenon atoms in the detector,
$a_{0}=1/(\alpha_{\rm{em}}m_{e})$ is the Bohr radius with $\alpha_{\rm{em}}=1/137$,
$F(q) \approx 1$ is the DM form factor,  
and $R_{s}$ is the resolution function which accounts for the ``efficiency" of the detector.
We will simply take the Gaussian distribution for the reconstructed energy for numerical analysis 
\begin{eqnarray}{\label{Rs}}
R_{s}(E, E_{e})=\frac{\alpha(E)}{\sqrt{2\pi}\sigma}\exp\left[-\frac{(E-E_{e})^{2}}{2\sigma^{2}}\right],
 \end{eqnarray}
where  $\alpha(E)$ is the efficiency \cite{Aprile:2020tmw} and $\sigma=a\sqrt{E_{e}}+bE_{e}$, with $a=(0.310\pm 0.004)$ $
\sqrt{\rm{keV}}$  and $b=0.0037\pm 0.0003$, respectively.  
\\
\\

%%%%%%%%%%%%%%%%%%%%%%%%%%%%%%%%%%%%%%%%%%%%%%%%
\begin{figure}
\centering
\includegraphics[width=9cm,height=9cm]{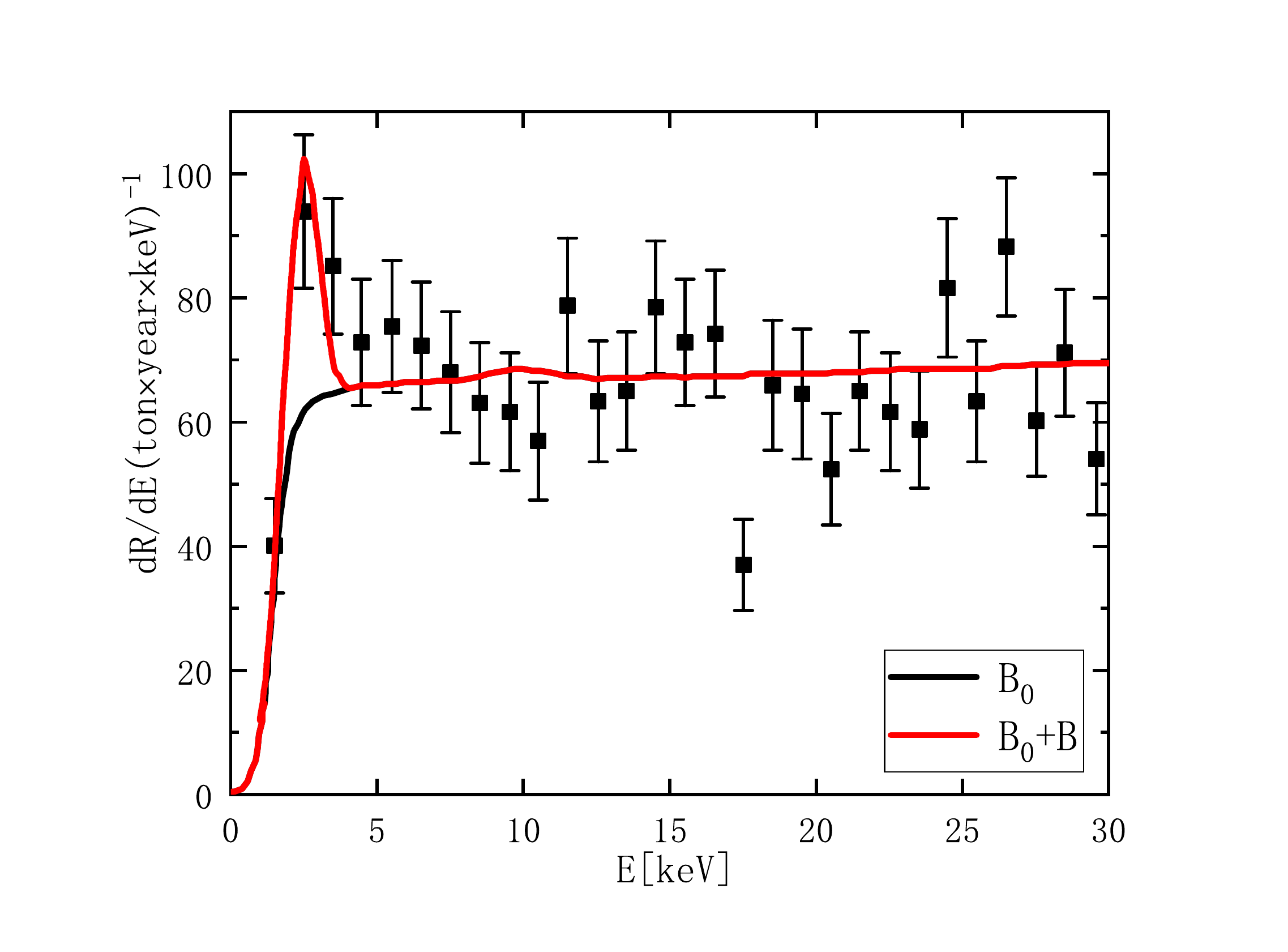}
\centering
\caption{Fit to the observed data \cite{Aprile:2020tmw} about the number of events $dR/dE$ $(\rm{ton~year~keV})^{-1}$ as function of the reconstructed electron recoil energy  
with five different sets of benchmark values C1 to C5 in Table.\ref{benchmark},
where $B_{0}$ and $B$ represents the background and the daughter DM contribution, respectively.}
\label{fit}
\end{figure}
%%%%%%%%%%%%%%%%%%%%%%%%%%%%%%%%%%%%%%%%%%%%%%%

%%%%%%%%%%%%%%%%%%%%%%%%%%%%%%%%%%%%%%%%%%%%%%%%
\begin{table}[htb!]
\begin{center}
\begin{tabular}{cccccc}
\hline\hline
 ~~~~&$m_A$ (MeV)~~ & $m_{B}$(MeV) ~~& $\upsilon_{B}/c$ ~~& $\tau_{A}/\rm{t}_{0}$ ~~& $\bar{\sigma}_{e}$ $(\rm{cm}^{2})$ \\ \hline
C1 & 10.05 & 5 &  0.1 & $3$ &  $ 4.36  \times 10^{-45}$ \\
C2 & 20.1&  10  &  0.1 & $3$ &  $9.46 \times 10^{-45}$  \\
C3 & 40.2  &  20 &  0.1 & $3$ &  $1.89 \times 10^{-44}$  \\
C4 &  100.5   &  50 & 0.1 & $3$ &  $4.22 \times 10^{-44}$  \\
C5 & 201  & 100 &  0.1 & $3$ &  $ 8.06  \times 10^{-44}$  \\
\hline \hline
\end{tabular}
\caption{Five sets of benchmark values which yield the same fit as shown in Fig.\ref{fit},
where the required values of $\bar{\sigma}_{e}$ can be understood as an output parameter.}
\label{benchmark}
\end{center}
\end{table}
%%%%%%%%%%%%%%%%%%%%%%%%%%%%%%%%%%%%%%%%%%%%%%%

Fig.\ref{fit} shows the fit to the reported XENON1T data \cite{Aprile:2020tmw} with 
five different sets of benchmark values C1 to C5 as explicitly shown in Table.\ref{benchmark}.
In individual case therein, we have chosen fixed value $\tau_{A}=3~\rm{t}_0$, 
under which $m_{A}\approx 2m_{B}$ take the mass ranges of $10-200 $ MeV and $\upsilon_{B}/c  = 0.1$.
The values of $\bar{\sigma}_{e}$ inferred from the observed XENON excess varies from $\mathcal{O}(10^{-45})$ cm$^{2}$ to  $\mathcal{O}(10^{-44})$ cm$^{2}$.

\section{Dark Matter Constraints}
%%% constraint on A and B
Now we turn to possible constraints on the dark matter particles $A$ and $B$.
Since the interaction in Eq.(\ref{decay}) yields too small annihilation cross section for $A$
to accommodate the required thermal annihilation cross section,
$A$ has to communicate either with the Standard Model (SM) sector e.g. via the same mediator as $B$, 
or mainly with other unstable freedoms in the dark sector. 
In the former situation, some constraints on $B$ as below can be placed on $A$ as well. 

With the communication between $B$ and the SM sector as inferred from the XENON1T excess, 
we can at least place the following constraints.
\begin{itemize}
\item The daughter DM B-free electron scattering cross section $\bar{\sigma}_e$, 
extracted from the XENON1T excess, 
can be used to constrain the model parameters.
Based on the measurements on $\bar{\sigma}_e$ within various electron recoil energies,  
 the light daughter DM can be probed either by the current XENON1T \cite{Aprile:2019xxb,Essig:2017kqs} or 
 the future SuperCDMS \cite{Essig:2015cda} experiments.\\
\item Similar to the DM B-electron scattering, 
we can also constrain $m_B$ from the annihilation cross section $\sigma^{B}_{\rm{ann}}(BB\rightarrow e^{+}e^{-})$,
based on the cross symmetry between the two Feynman diagrams related to these two processes.
While experiments such as AMS-01 \cite{Aguilar:2007yf},  AMS-02 \cite{Aguilar:2013qda} or PAMELA \cite{Adriani:2013uda} have not yet placed viable bounds on $\sigma^{B}_{\rm{ann}}$ in the sub GeV-scale DM mass, 
the Planck data \cite{Ade:2015xua} is able to constrain $m_B$ down to $\sim 1$ MeV.\\
\item Lastly, the coupling of the mediator to electron can be constrained by colliders such as BaBar, LEP and LHC.
\end{itemize}

\begin{figure}[htb!]
\centering
\includegraphics[width=9cm,height=9cm]{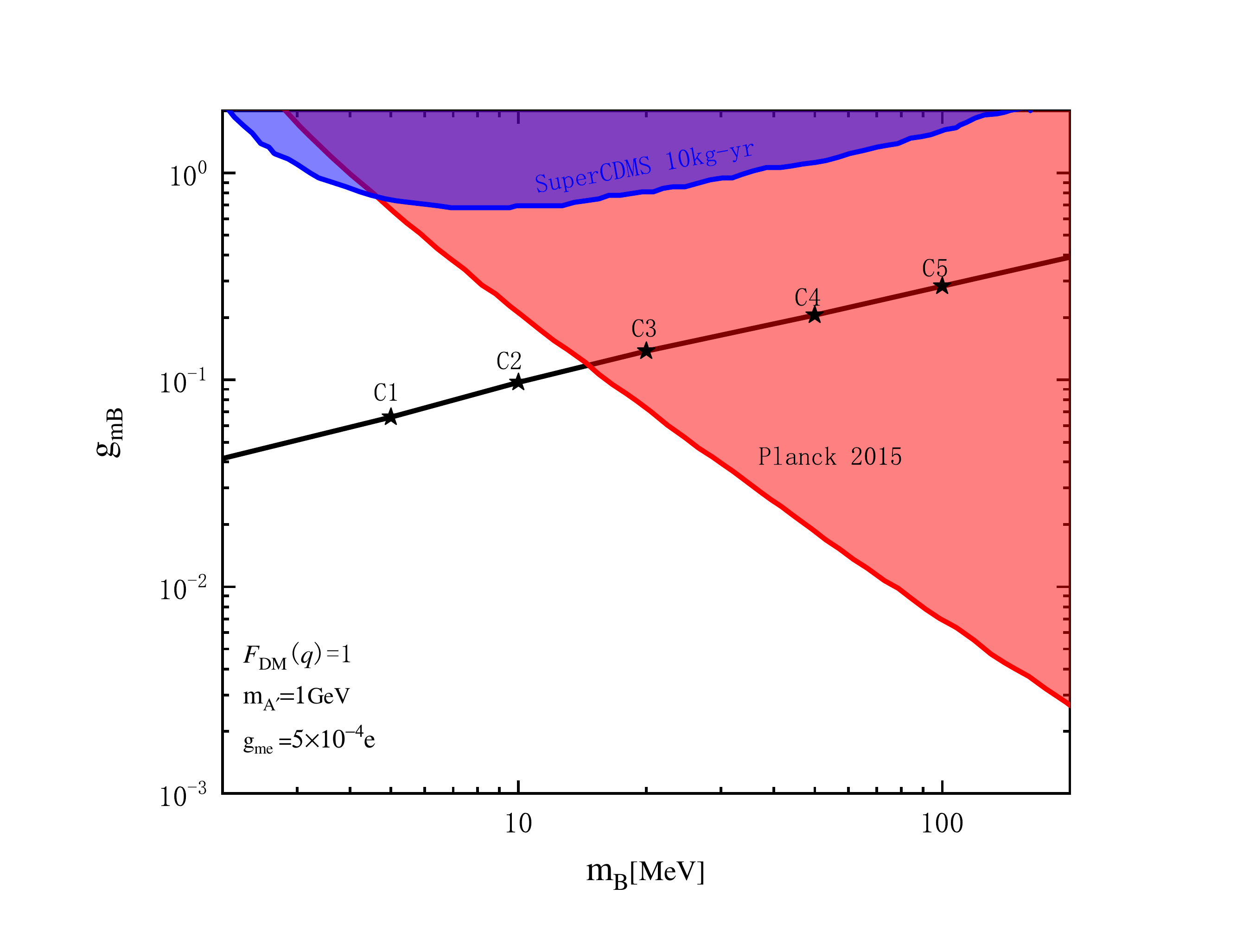}
\centering
\caption{Constraints on the parameter space (in black curve) in the dark photon model with $m_{A'}=1$ GeV, $g_{me}=5\times 10^{-4} e$ and the DM form factor $F(q)=1$. 
We have simultaneously shown the SuperCDMS limit \cite{Essig:2015cda} (in blue) and the Planck 2015 limit \cite{Slatyer:2015jla} (in red), where the shaded regions are excluded.}
\label{DMe}
\end{figure}

For illustration, we show in Fig.\ref{DMe} the constraints in the specific dark photon model with the mediator identified as 
a new vector boson $A'$, where $m_{A'}=1$ GeV and $g_{me}=5\times 10^{-4} e$ have been adopted in the light of BaBar
data \cite{Lees:2014xha}.
In this figure, we have simultaneously shown the SuperCDMS limit \cite{Essig:2015cda} (in blue) without relativistic effect \cite{Pandey:2018esq}, the Planck 2015 limit \cite{Slatyer:2015jla} (in red) and the parameter space (in black curve) together with the benchmark values in Table.\ref{benchmark}.
Relatively weaker XENON1T limit has been ignored.
One observes that in this explicit model $m_B$ beneath $\sim 15$ MeV survives.

%%%%%%%%%%%%%%%%%%%%%%%%%%%%%%%%%%%%%%%%%%%%%%%%
%%%%%%%%%%%%%%%%%%%%%%%%%%%%%%%%%%%%%%%%%%%%%%%

Compared to the benchmark values in Table.\ref{benchmark}, 
given fixed $m_B$ one can obtain larger $\bar{\sigma}_{e}$ or alternatively larger $g_{mB}g_{me}$ by taking larger $\tau_{A}$,
since they are linearly correlated to each other in $dR/dE \sim  \bar{\sigma}_{e}(\rm{t}_{0}/\tau_{A})$ for $\rm{t}_{0}\ll \tau_{A}$ in Eq.(\ref{R}).
However, an increase of $\tau_A$ will simultaneously lead to linearly enhanced experimental limits in Fig.\ref{DMe}.
These trends together imply that adjusting $\tau_A$  is unable to alter the SuperCDMS sensitivity as illustrated in Fig.\ref{DMe}.

\section{Cosmological Constraints}
The decaying DM model are constrained both by the CMB and the LSS experiments for a varying dark matter energy density with time. 
In our scenario, it reads from Eq.(\ref{number}) 
\begin{eqnarray}{\label{density}}
\rho_{\rm{dm}}(t)=\rho_{\rm{dm},0}\left[e^{-t/\tau_{A}}+\frac{2m_B}{m_{A}}\left(1-e^{-t/\tau_{A}}\right)\right]a^{-3}(t),\nonumber\\
 \end{eqnarray}
 Compared to the baseline $\Lambda$CDM cosmology, 
the DM relic density in Eq.(\ref{density}) is altered by a magnitude of order $\mid\Delta \rho_{\rm{dm}}/\rho_{\rm{dm},0}\mid\approx (1-\frac{2m_{B}}{m_{A}})t/\tau_{A}<10^{-3}$ in the small redshift region for the benchmark values in Table.\ref{benchmark}, as a result of highly degenerate dark matter mass relation $m_{A}\approx 2m_{B}$ required by the XENON excess.

For the CMB experiment \cite{Aghanim:2018eyx},
it mainly affects the temperature power spectrum $C_{TT}$ in terms of the integrated Sachs-Wolfe  effect,
which relies on the cosmological evolution of Universe after the last scattering. 
Due to the small fraction in $\rho_{\rm{dm}}$ given by Eq.(\ref{density}) relative to what attempts to explain the Hubble tension \cite{Vattis:2019efj,Anchordoqui:2015lqa,Bringmann:2018jpr,Clark:2020miy}, 
the effect on $C_{TT}$ in our scenario is negligible.
For the LSS experiments, the DM power spectrum $\delta=\delta\rho_{\rm{dm}}/\rho_{\rm{dm}}$ evolves with time as
\begin{eqnarray}{\label{matter}}
\ddot{\delta}+2H\dot{\delta}-4\pi G \rho_{\rm{dm}}\delta=0,
 \end{eqnarray}
where $G$ is the Newton's constant and $H$ is the Hubble rate.
The small fraction in $\rho_{\rm{dm}}$ gives rise to a fraction in the DM power spectrum $\delta$ less than the order of a percent level, 
which is far beyond the reach of future LSS experiments such as the Dark Energy Spectroscopic Instrument \cite{Aghamousa:2016zmz}.

\section{Conclusions}
In this study we have proposed a novel decaying cold DM scenario 
in which the cold parent DM $A$ decays to the daughter particle $B$, 
with the lifetime $\tau_{A}$ larger than the age of Universe.
Firstly we have shown that in this scenario the observed excess of the electron recoil at the XENON1T in the energy range $2-3$ keV can be addressed 
by the daughter DM B-electron elastic scattering with the DM mass ranges $m_{A}\approx 2m_{B}\sim 10-200$ MeV.
Moreover, we have verified that because of small DM B-electron scattering cross section this model is consistent with  limits both from the DM direct and indirect detections,
while as a result of suppression on the magnitude of the fraction in the DM energy density due to the highly degenerate mass relation imposed by the XENON excess, 
this model does not violate either the CMB measurements on the temperature power spectrum or the LSS constraints on the DM power spectrum.
Finally, there are a few directions in our DM scenario which deserve further investigation.  
Especially, if we are allowed to adopt the lifetime of the parent DM obviously smaller than the age of Universe,
it is not unlikely to resolve the XENON1T excess and the Hubble tension simultaneously with a decaying cold DM.

\section*{Acknowledgments}
The research is supported in part by the National Natural Science Foundation of China with Grant No. 11775039
and the Fundamental Research Funds for the Central Universities at Chongqing University with Grant No. cqu2017hbrc1B05.

\end{document}